\documentstyle[12pt]{article}

\begin{document}

\newcommand{\eq}[1]{eq.~(\ref{#1})}
\newcommand{\eqs}[2]{eqs.(\ref{#1}, \ref{#2})}
\newcommand{\eqss}[3]{eqs.(\ref{#1}, \ref{#2}, ref{#3})}
\newcommand{\ur}[1]{~(\ref{#1})}
\newcommand{\urs}[2]{~(\ref{#1},~\ref{#2})}
\newcommand{\urss}[3]{~(\ref{#1},~\ref{#2},~\ref{#3})}
\newcommand{\Eq}[1]{Eq.~(\ref{#1})}
\newcommand{\Eqs}[2]{Eqs.(\ref{#1},\ref{#2})}
\newcommand{\fig}[1]{Fig.~\ref{#1}}
\newcommand{\figs}[2]{Figs.\ref{#1},\ref{#2}}
\newcommand{\figss}[3]{Figs.\ref{#1},\ref{#2},\ref{#3}}
\newcommand{\beq}{\begin{equation}}
\newcommand{\eeq}{\end{equation}}
\newcommand{\e}{\varepsilon}
\newcommand{\ee}{\epsilon}
\newcommand{\bI}{\bar I}
\newcommand{\eoe}{(\bar{\eta}O\eta)}
\newcommand{\ui}{U_{int}(R,\rho_{1,2},O_{1,2})}
\newcommand{\thb}[3]{\bar{\eta}^{#1}_{#2 #3}}
\newcommand{\tha}[3]{\eta^{#1}_{#2 #3}}
\newcommand{\la}[1]{\label{#1}}
\newcommand{\YM}{Yang--Mills~}
\newcommand{\N}{$N_{CS}\;$}
\newcommand{\Nz}{$N_{CS}=0\;$}
\newcommand{\No}{$N_{CS}=1\;$}
\newcommand{\x}{{\bf x}}
\newcommand{\r}{{\bf r}}
\newcommand{\IIs}{$I$'s and ${\bar I}$'s }
\newcommand{\II}{$I{\bar I}$ }

\vspace{1.5cm}
\begin{center} \Large
CHIRAL SYMMETRY BREAKING BY INSTANTONS \footnote{Lectures at
the Enrico Fermi School in Physics, Varenna, June 27 -- July 7, 1995}
\end{center}
\vspace{1.2cm}

\begin{center} {\large Dmitri Diakonov} \\
{\small\it Petersburg Nuclear Physics Institute, Gatchina, St.Petersburg
188350, Russia} \end{center}

\vspace{2cm}

\section{Foreword}
\setcounter{equation}{0}
\def\theequation{\arabic{section}.\arabic{equation}}

Instantons are certain configurations of the \YM potentials
$A_\mu^a(x)$ satisfying the equations of motion $D_\mu^{ab}
F_{\mu\nu}^b = 0$ in euclidean space, i.e. in imaginary time. The
solution has been found by Belavin, Polyakov, Schwartz and Tiupkin in
1975 \cite{BPST}; the name "instanton" has been suggested by 't Hooft
in 1976 \cite{tH}, who also made a major contribution to the
investigation of the instantons properties.

In QCD instantons are the best studied non-perturbative effects,
leading to the formation of the gluon condensate \cite{SVZ} and of the
so-called topological susceptibility needed to cure the $U(1)$ paradox
\cite{tH,Akas}. The QCD instanton vacuum has been studied starting from
the pioneering works in the end of the seventies \cite{CDG,Sh1}; a
quantitative treatment of the instanton ensemble, based on the Feynman
variational principle, has been developed in ref. \cite{DP1}. The most
striking success of the QCD instanton vacuum is its capacity to provide
a beautiful mechanism of the spontaneous chiral symmetry breaking
\cite{DP2,DP3,Pob}. Moreover, the instanton vacuum leads to a very
reasonable effective chiral lagrangian at low energies, including the
Wess--Zumino term, etc., which, in its turn gives a nice description of
nucleons as chiral quark solitons \cite{DPP2}.

It should be stressed that literally speaking instantons do not lead
to confinement, though they induce a growing potential for heavy
quarks at intermediate separations \cite{DPP1}; asymptotically it
flattens out \cite{CDG,Sh1a}. However, it has been realized a decade ago
\cite{DP2,DP3,D}, that it is chiral symmetry breaking and not
confinement that determines the basic characteristics of {\em light}
hadrons (one would probably need an explicit confinement to get
the properties of highly excited hadrons). Therefore, since the
instanton vacuum describes well the physics of the chiral symmetry
breaking, one would expect that instantons do explain the
properties of light hadrons, both mesons and baryons. Indeed,
a detailed numerical study of dozens of correlation functions in the
instanton medium undertaken by Shuryak, Verbaarschot and Sch\"{a}fer
\cite{Sh2} (earlier certain correlation functions were computed
analytically in refs. \cite{DP2,DP3}) demonstrated an impressing
agreement with the phenomenology \cite{Sh3} and, more recently, with
direct lattice measurements \cite{CGHN}.  As to baryons, the
instanton-motivated chiral quark soliton model \cite{DPP2} also leads
to a very reasonable description of dozens of baryon characteristics
(for a review see ref.\cite{G}).

More recently the instanton vacuum was studied in direct lattice
experiments by the so-called cooling procedure
\cite{T,ILMSS,PV,CDMPV,CGHN,MS}, see also the proceedings of
the Lattice-94 meeting for a more complete list of references. It was
demonstrated that instantons and antiinstantons ($I$'s and ${\bar I}$'s
for short) are the only non-perturbative gluon configurations surviving
after a sufficient smearing of the quantum gluon fluctuations. The
measured properties of the $I{\bar I}$ medium appeared
\cite{PV,CGHN,MS} to be close to that computed from the variational
principle \cite{DP1} and to what had been suggested by Shuryak in the
beginning of the 80's \cite{Sh1} from phenomenological considerations.

Cooling down the quantum fluctuations above instantons kills both the
one-gluon exchange and the linear confining potential (a small residual
string tension observed in the cooled vacuum \cite{PV,CGHN} is probably
due to the instanton-induced rising potential at intermediate distances
\cite{DPP1}). Nevertheless, in the cooled vacuum where only \IIs are left,
the correlation functions of various mesonic and baryonic currents, as
well as the density-density correlation functions appear to be quite
similar to those of the true or "hot" vacuum \cite{CGHN}.  We consider
it to be a remarkable confirmation of the ideology which has been put
forward quite some time ago \cite{Sh1,DP2,DP3,D}: instantons are
responsible for the basic properties of the QCD vacuum and of the
low-mass hadrons, while the confinement, fundamental as it is, does
not readily manifest itself in those properties.

These lectures are aimed as an introduction to understanding the above
works.

\section{Periodicity of the Yang--Mills potential energy}
\setcounter{equation}{0}
\def\theequation{\arabic{section}.\arabic{equation}}

I start by explaining the physical meaning of instantons as
classical tunneling trajectories in imaginary (euclidean) time. To that
end I shall temporarily  work in the $A_0^a=0$ gauge, called
Weyl or Hamiltonian gauge, and forget about fermions for a while. The
remaining pure \YM or "pure glue" theory is nonetheless
non-trivial, since gluons are self-interacting. For simplicity I start
from the $SU(2)$ gauge group.

The spatial YM potentials $A_i^a(\x, t)$ can be considered
as an infinite set of the coordinates of the system, where $i=1,2,3,\;\;
a=1,2,3$ and $\x$ are "labels" denoting various coordinates. The
YM action is

\beq
S=\frac{1}{4g^2}\int\! d^4x\; F_{\mu\nu}^aF_{\mu\nu}^a=\int dt
\left(\frac{1}{2g^2}\int\! d^3\x\;{\bf E}^2 - \frac{1}{2g^2}\int\!d^3\x
\;{\bf B}^2 \right)
\la{YMA}\eeq
where ${\bf E}$ is the electric field stregth,

\beq
E_i^a(\x,t)={\dot A}_i^a(\x,t)
\la{E}\eeq
(the dot stands for the time derivative), and ${\bf B}$ is the magnetic
field stregth,

\beq
B_i^a(\x,t)=\frac{1}{2}\epsilon_{ijk}\left(\partial_jA_k^a-
\partial_kA_j^a+\epsilon^{abc}A_j^bA_k^c\right).
\la{B}\eeq

Apparently, the first tirm in \eq{YMA} is the kinetic energy of the
system of coordinates $\{A_i^a(\x,t)\}$ while the second term is minus
the potential energy being just the magnetic energy of the field.
The simple and transparent form of \eq{E} is the advantage of the Weyl
gauge.

Let us introduce an important quantity called the Pontryagin index or
the four-dimensional topological charge of the YM fields:

\beq
Q_T=\frac{1}{32\pi^2}\int d^4x\; F_{\mu\nu}^a {\tilde F}_{\mu\nu}^a,
\;\;\;\;{\tilde F}_{\mu\nu}^a \equiv \frac{1}{2}
\epsilon_{\mu\nu\alpha\beta}F_{\alpha\beta}^a.
\la{FFd}\eeq
The integrand in \eq{FFd} is a full derivative of a four-vector $K_\mu$:

\beq
\frac{1}{32\pi^2} F_{\mu\nu}^a {\tilde F}_{\mu\nu}^a =\partial_\mu
K_\mu,\;\;\;
K_\mu=\frac{1}{16\pi^2}\epsilon_{\mu\alpha\beta\gamma}
\left(A_\alpha^a\partial_\beta A_\gamma^a+\frac{1}{3}\epsilon^{abc}
A_\alpha^a A_\beta^b A_\gamma^c\right).
\la{K}\eeq

Therefore, assuming the fields $A_\mu$ are decreasing rapidly enough at
spatial infinity, one can rewrite the 4-dimensional topological charge
\ur{FFd} as

\beq
Q_T=\int d^4x (\partial_0 K_0-\partial_i K_i)=\int dt \frac{d}{dt}
\int d^3\x K_0.
\la{Gauss}\eeq
Introducing the {\it Chern--Simons number}

\beq
N_{CS}=\int d^3\x\;K_0=\frac{1}{16\pi^2}\int d^3\x\:\epsilon_{ijk}
\left(A_i^a\partial_j A_k^a+\frac{1}{3}\epsilon^{abc}
A_i^a A_j^b A_k^c\right)
\la{NCS}\eeq
we see from \eq{Gauss} that $Q_T$ can be rewritten as the difference
of the Chern--Simons numbers characterizing the fields at $t=\pm\infty$:

\beq
Q_T=N_{CS}(+\infty)-N_{CS}(-\infty).
\la{dif}\eeq

The Chern--Simons characteristics of the field has an important property
that it can change by integers under large gauge transformations. Indeed,
under a general time-independent gauge transformation,

\beq
A_i \rightarrow U^\dagger A_iU + iU^\dagger\partial_iU, \;\;\;
A_i\equiv A_i^a\frac{\tau^a}{2},
\la{gt}\eeq
the Chern--Simons number transforms as follows:

\beq
N_{CS}\rightarrow N_{CS}+N_W
\la{CSt}\eeq
where $N_W$ is the winding number of the gauge transformation \ur{gt}:

\beq
N_W=\frac{1}{24\pi^2}\int d^3\x\; \epsilon_{ijk}\left[(U^\dagger\partial_iU)
(U^\dagger\partial_jU) (U^\dagger\partial_kU)\right].
\la{wn}\eeq

The $SU(2)$ unitary matrix $U$ of the gauge
transformation \ur{gt} can be viewed as a mapping from the 3-dimensional
space onto the 3-dimensional sphere of parameters $S_3$. If at spatial
infinity we wish to have the same matrix $U$ independently of the way we
approach the infinity (and this is what is usually assumed), then the
spatial infinity is in fact one point, so the mapping is topologically
equivalent to that from $S_3$ to $S_3$. This mapping is known to be
non-trivial, meaning that mappings with different winding numbers
are irreducible by smooth transformations to one another. The winding
number of the gauge transformation is, analytically, given by \eq{wn}.
As it is common for topological characteristics, the integrand in \ur{wn}
is in fact a full derivative. For example, if we take the matrix $U(\x)$
in a "hedgehog" form, $U=\exp[i(\r{\bf \tau}/r\cdot P(r)]$, \eq{wn} can be
rewritten as

\beq
N_W=\frac{2}{\pi}\int dr \sin^2 P \frac{dP}{dr} = \frac{1}{\pi}
\left[P-\frac{\sin 2P}{2}\right]_0^\infty = \mbox{integer}
\la{wnh}\eeq
since $P(r)$ both at zero and at infinity needs to be multiples of $\pi$
if we wish $U(\r)$ to be unambigiously defined in the origin and
at the infinity.

Let us return now to the potential energy of the YM fields,

\beq
{\cal V}=\frac{1}{2g^2} \int d^3\x \left(B_i^a\right)^2.
\la{potene}\eeq

One can imagine plotting the potential energy surfaces over the
Hilbert space of the coordinates $A_i^a(\x)$. It will be some complicated
mountain country, like around Varenna. If the field happens to be a
pure gauge, $A_i= iU^\dagger\partial_i U$, the potential energy  at
such points of the Hilbert space is naturally zero. Imagine that we
move along the "generalized coordinate" being the Chern--Simons number
\ur{NCS}, fixing all other coordinates whatever they are. Let us take
some point $A_i^a(\x)$ with the potential energy ${\cal V}$. If we move
to another point which is a gauge transformation of $A_i^a(\x)$ with a
winding number $N_W$, its potential energy will be exactly the same as
it is strictly gauge invariant.  However the Chern--Simons "coordinate"
of the new point will be shifted by an integer $N_W$ from the original
one. We arrive to the conclusion first pointed out by Faddeev
\cite{Fad} and Jackiw and Rebbi \cite{JR} in 1976, that the potential
energy of the YM fields is {\em periodic} in the particular coordinate
called the Chern--Simons number.

One may wish to plot this periodic dependence of the YM potential energy
on \N. Putting such a problem in a situation where the potential energy
depends also on an infinite number of other "coordinates", we imply that
one is to find the minimal energy path, say, from \Nz to \No, for
a given value of \N. However, the pure YM theory is scale-invariant
at the classical level, so to solve the problem one has to fix the
spatial size of the $A_i$ fields somehow. The situation is different in
the electro-weak theory where the scale invariance is explicitly broken
at the classical level by the non-zero Higgs vacuum expectation value.
Therefore the problem of finding the minimal energy pass in the mountain
country of the YM Hilbert space is well defined. It was solved by
Akiba, Kikuchi and Yanagida in 1988 \cite{AKY}, and the reader may
satisfy his or her curiosity and have a view of the periodic dependence
of the potential energy on the Chern--Simons number in that work (see
also ref.\cite{DPSSG} where this dependence is generalized to non-zero
matter density).

The static configuration of the $A_i$ fields corresponding to the saddle
point in the minimal-energy path and having exactly \N $=1/2$ is called
{\em sphaleron}. It was found many years ago by Dashen, Hasslacher and
Neveu \cite{DHN} and rediscovered in the context of the electro-weak
theory by Manton and Klinkhammer \cite{KM}. The sphaleron is the
"coordinate" of the top of the potential barrier separating two
zero-potential points, \Nz and \No. In the electro-weak theory
the height of the barrier is of the order of $m_W/\alpha$ ($m_W$ is
the mass of the $W$ boson and $\alpha$ is the gauge coupling constant).
In an unbroken YM theory like QCD the classical energy barrier between
topologically distinct vacua can be made infinitely small due to the
scale invariance. However that does not mean that the barriers are
easily penetrable: we shall calculate the transition amplitudes from
one mimimum to another in the next section.

\section{Instanton configurations}
\setcounter{equation}{0}
\def\theequation{\arabic{section}.\arabic{equation}}

In perturbation theory one deals with zero-point quantum-mechanical
fluctuations of the YM fields near one of the minima, say, at \Nz.
The non-linearity of the YM theory is taken into account as a
perturbation, and results in series in $g^2$ where $g$ is the gauge
coupling.  In that approach one is apparently missing a possibility for
the system to tunnel to another minimum, say, at \No. The tunneling is
a typical non-perturbative effect in the coupling constant, and
instantons have direct relation to the tunneling.

The tunneling amplitude can be estimated as $exp(-S)$, where $S$ is the
action along the classical trajectory in imaginary time, leading from
the minimum at \Nz at $t=-\infty$ to that at \No at $t=+\infty$
\cite{LL}. According to \eq{dif} the 4-dimensional topological charge
of such trajectory is $Q_T=1$. To find the best tunneling trajectory
having the largest amplitude one has thus to minimize the YM action
\ur{YMA} provided the topological charge \ur{FFd} is fixed to be unity.
This can be done using the following trick \cite{BPST}. Consider
the inequality

\[
0\leq \int d^4x \left(F_{\mu\nu}^a-{\tilde F}_{\mu\nu}^a\right)^2
\]
\beq
=\int d^4 x \left(2 F^2-2F{\tilde F}\right) =
8g^2S-64\pi^2Q_T,
\la{inequ}\eeq
hence the action is restricted from below:

\beq
S\geq \frac{8\pi^2}{g^2}Q_T =  \frac{8\pi^2}{g^2}.
\la{inequa}\eeq

Therefore, the minimal action for a trajectory with a unity topological
charge is equal to $8\pi^2/g^2$, which is achieved if the trajectory
satisfies the {\em self-duality} equation:

\beq
F_{\mu\nu}^a={\tilde F}_{\mu\nu}^a.
\la{selfdual}\eeq

Notice that any solution of \eq{selfdual} is simultaneously a solution of
the general YM equation of motion $D_\mu^{ab}F_{\mu\nu}^b=0$: that is
because the "second pair" of the Maxwell equations,
$D_\mu^{ab}{\tilde F}_{\mu\nu}^b=0$, is satisfied identically.

To solve \eq{selfdual} let us recall a few facts about the Lorentz group
$SO(3,1)$. Since we a talking about the tunneling fields which can only
develop in imaginary time, it means that we have to consider the fields
in euclidean space-time, so that the Lorentz group is just $SO(4) =
SU(2)\times SU(2)$. The gauge potentials $A_\mu$ belong to the
$(\frac{1}{2},\frac{1}{2})$ representation of the $SU(2)\times SU(2)$
group, while the field strength $F_{\mu\nu}$ belongs to the $(1,1)$
representation. In other words it means that one linear combination
of $F_{\mu\nu}$ transforms as a vector of the left $SU(2)$, and another
combination transforms as a vector of the right $SU(2)$. These
combinations are

\beq
F_L^A=\eta_{\mu\nu}^A(F_{\mu\nu}+\tilde{F}_{\mu\nu}),\;\;\;\;
F_R^A=\bar{\eta}_{\mu\nu}^A(F_{\mu\nu}-\tilde{F}_{\mu\nu}),
\la{oneone}\eeq
where $\eta, \bar{\eta}$ are the so-called 't Hooft symbols described
in ref.\cite{tH}, see also below. We see therefore that a self-dual
field strength is a vector of the left $SU(2)$ while its right part is
zero.  Keeping that experience in mind we look for the solution of the
self-dual equation in the form

\beq
A_\mu^a=\thb{\mu}{\nu}{a} x_\nu\frac{1+\Phi(x^2)}{x^2}.
\la{tHanz}\eeq

Using the formulae for the $\eta$ symbols from ref.\cite{tH} one can
easily check that the YM action can be then rewritten as

\beq
S=\frac{8\pi^2}{g^2}\frac{3}{2} \int d\tau
\left[\frac{1}{2}\left(\frac{d\Phi}{d\tau}\right)^2+
\frac{1}{8}(\Phi^2-1)^2\right], \;\;\;\;
\tau = \ln\left(\frac{x^2}{\rho^2}\right).
\la{doublewell}\eeq

This can be recognized as the action of the double-well potential whose
minima lie at $\Phi=\pm 1$, and $\tau$ plays the role of "time", $\rho$
is an arbitrary scale. The trajectory which tunnels from $1$ at
$\tau=-\infty$ to $-1$ at $\tau = +\infty$ is

\beq
\Phi=-\tanh\left(\frac{\tau}{2}\right),
\la{QMinst}\eeq
and its action \ur{doublewell} is $S=8\pi^2/g^2$, as needed. Substituting
the solution \ur{QMinst} into \ur{tHanz} we get

\beq
A_\mu^a(x)=\frac{2\thb{\mu}{\nu}{a}\rho^2}{x^2(x^2+\rho^2)}.
\la{YMinst}\eeq

The correspondent field strength is

\beq
F_{\mu\nu}^a=-\frac{4\rho^2}{(x^2+\rho^2)^2}\left(\thb{\mu}{\nu}{a}-
2\thb{\mu}{\alpha}{a}\frac{x_\alpha x_\nu}{x^2}-
2\thb{\beta}{\nu}{a}\frac{x_\mu x_\beta}{x^2}\right),\;\;\;
F_{\mu\nu}^aF_{\mu\nu}^a=\frac{192\rho^4}{(x^2+\rho^2)^4},
\la{fstrength}\eeq
and satisfies the self-duality condition \ur{selfdual}.

The {\em anti-instanton} corresponding to tunneling in the opposite
direction, from \No to \Nz, satisfies the {\em anti}-self-dual
equation, with $\tilde{F}\rightarrow -\tilde{F}$; its concrete form is
given by \eqs{YMinst}{fstrength} with the replacement
$\bar{\eta}\rightarrow \eta$.

\Eqs{YMinst}{fstrength} describe the field of the instanton in the
singular Lorentz gauge; the singularity of $A_\mu$ at $x^2=0$ is a gauge
artifact: the gauge-invariant field strength squared is smooth at the
origin. The formulae for instantons are more simple in the Lorentz gauge,
and I shall use it further on.

The instanton field, \eq{YMinst}, depends on an arbitrary scale
parameter $\rho$ which we shall call the instanton size, while the
action, being scale invariant, is independent of $\rho$. One can
obviously shift the position of the instanton to an arbitrary 4-point
$z_\mu$ -- the action will not change either. Finally, one can rotate
the instanton field in colour space by constant unitary matrices $U$.
For the $SU(2)$ gauge group this rotation is characterized by 3
parameters, {\it e.g.} by Euler angles. For a general $SU(N_c)$ group
the number of parameters is $N_c^2-1$ (the total number of the
$SU(N_c)$ generators) {\em minus} $(N_c-2)^2$ (the number of generators
which do not effect the left upper $2\times 2$ corner where the
standard $SU(2)$ instanton \ur{YMinst} is residing), that is $4N_c-5$.
These degrees of freedom are called instanton orientation in colour
space.  All in all there are

\beq
4\; {\mbox (centre)}\;\;+\;\;1\; {\mbox (size)}\;\;+\;\;
(4N_c-5)\; {\mbox (orientations)}\;\;=\;\;4N_c
\la{collcoo}\eeq
so-called collective coordinates desribing the field of the instanton,
of which the action is independent.

It is convenient to indroduce $2\times 2$ matrices

\beq
\sigma^{\pm}_\mu = (\pm i \overrightarrow{\sigma}, 1),\;\;\;
x^{\pm}=x_\mu\sigma^{\pm}_\mu,
\la{sigma}\eeq
such that

\beq
2i\tau^a
\tha{\mu}{\nu}{a}=\sigma^+_\mu\sigma^-_\nu
-\sigma^+_\nu\sigma^-_\mu; \;\;\;
2i\tau^a
\thb{\mu}{\nu}{a}=\sigma^-_\mu\sigma^+_\nu
-\sigma^-_\nu\sigma^+_\mu,
\la{thsig}\eeq
then the instanton field with arbitrary center $z_\mu$, size $\rho$ and
colour orientation $U$ in the $SU(N_c)$ gauge group can be written as

\beq
A_\mu=A_\mu^at^a
=\frac{-i\rho^2U[\sigma^-_\mu(x-z)^+-(x-z)_\mu]U^\dagger}
{(x-z)^2[\rho^2+(x-z)^2]},\;\;\;Tr(t^at^b)=\frac{1}{2}\delta^{ab},
\la{instgen}\eeq
or

\beq
A_\mu^a=\frac{2\rho^2O^{ab}\thb{\mu}{\nu}{b}(x-z)_\mu}
{(x-z)^2[\rho^2+(x-z)^2]},\;\;\;
O^{ab}=Tr(U^\dagger t^aU\sigma^b),\;\;\;O^{ab}O^{ac}=\delta^{bc}.
\la{instgena}\eeq

Physically, one can think of instantons in two ways: on one hand it is
a tunneling {\em process} occuring in imaginary time (this
interpretation belongs to V.Gribov, 1976), on the other hand it is a
localized {\em pseudoparticle} in the euclidean space (A.Polyakov,
1977 \cite{Pol}).

\section{Gluon condensate}
\setcounter{equation}{0}
\def\theequation{\arabic{section}.\arabic{equation}}

The QCD perturbation theory implies that the fields $A_i^a(\x)$ are
performing quantum zero-point oscillations; in the lowest order these
are just plane waves with arbitrary frequences. The aggregate energy of
these zero-point oscillations, $({\bf B}^2+{\bf E}^2)/2$, is divergent
as the fourth power of the cutoff frequency, however for any state one
has $\langle F_{\mu\nu}^2\rangle = 2\langle{\bf B}^2-{\bf E}^2\rangle =
0$, which is just a manifestation of the virial theorem for harmonic
oscillators:  the average potential energy is equal to that of the
kinetic (I am temporarily in the Minkowski space). One can prove
that this is also true in any order of the perturbation theory in the
coupling constant, provided one does not violate the Lorentz symmetry
and the renormalization properties of the theory. Meanwhile, we know
from the QCD sum rules phenomenology that the QCD vacuum posseses what
is called {\em gluon condensate} \cite{SVZ}:

\beq
\frac{1}{32\pi^2}\langle F_{\mu\nu}^aF_{\mu\nu}^a\rangle
\simeq (200\; MeV)^4 \;\;>\;0.
\la{glcond}\eeq

Instantons suggest an immediate explanation of this basic property of
QCD. Indeed, instanton is a tunneling process, it occurs in imaginary
time; therefore in Minkowski space one has $E_i^a=\pm iB_i^a$ (this is
actually the duality \eq{selfdual}). Therefore, one gets a chance to
explain the gluon condensate. In euclidean space the electric field is
real as well as the magnetic one, and the gluon condensate is just the
average action density. Let us make a quick estimate of its value.

Let the total number of \IIs in the 4-dimensional volume $V$ be $N$.
Assuming that the average separations of instantons are larger than
their average sizes (to be justified below), we can estimate the total
action of the ensemble as the sum of invidual actions (see
\eq{inequa}):

\beq
\langle F_{\mu\nu}^2\rangle V =\int d^4x F_{\mu\nu}^2
\simeq N\cdot 32\pi^2,
\la{totact}\eeq
hence the gluon condensate is directly related to the instanton density
in the 4-dimensional euclidean space-time:

\beq
\frac{1}{32\pi^2}\langle F_{\mu\nu}^aF_{\mu\nu}^a\rangle
\simeq \frac{N}{V}\equiv \frac{1}{{\bar R}^4}.
\la{glcondest}\eeq
In order to get the phenomenological value of the condensate one needs
thus to have the average separation between pseudoparticles \cite{Sh1}

\beq
{\bar R}\simeq\frac{1}{200\;MeV}=1\;fm.
\la{avsep}\eeq

There is another point of view on the gluon condensate which I describe
briefly. In principle, all information about field theory is contained
in the partition function being the functional integral over the
fields. In the euclidean formulation it is

\beq
{\cal Z}=\int DA_\mu exp\left(-\frac{1}{4g^2}\int d^4x
F_{\mu\nu}^2\right) \stackrel{T\rightarrow\infty}{\longrightarrow}
e^{-ET},
\la{partfu}\eeq
where I have used that at large (euclidean) time $T$ the partition
function picks up the ground state or vacuum energy $E$.
For the sake of brevity I do not write the gauge fixing and
Faddeev--Popov ghost terms. If the
state is homogeneous, the energy can be written as
$E=\theta_{44}V^{(3)}$ where $\theta_{\mu\nu}$ is the stress-energy
tensor and $V^{(3)}$ is the 3-volume of the system.
Hence, at large 4-volumes $V=V^{(3)}T$ the partition function is
${\cal Z}=\exp(-\theta_{44}V)$. This $\theta_{44}$ includes zero-point
oscillations and diverges badly. A more reasonable quantity is the
partition function, normalized to the partition function understood as
a perturbative expansion about the zero-field vacuum\footnote{The
latter can be distinguished from the former by imposing a condition
that it does not contain integration over singular Yang--Mills
potentials; recall that the instanton potentials are singular at the
origins.},

\beq
\frac{{\cal Z}}{{\cal Z}_{P.T.}}= \exp\left[-(\theta_{44}
-\theta_{44}^{P.T.})V\right].
\la{pfn}\eeq

We expect that the non-perturbative vacuum energy density
$\theta_{44}-\theta_{44}^{P.T.}$ is a negative quantity since we
have allowed for tunneling: as usual in quantum mechanics, it
lowers the ground state energy. If the vacuum is isotropical, one has
$\theta_{44}=\theta_{\mu\mu}/4$. Using the trace anomaly,

\beq
\theta_{\mu\mu}=\frac{\beta(g^2)}{4g^4}
\frac{\left(F_{\mu\nu}^a\right)^2}{32\pi^2}\simeq
-b\frac{F_{\mu\nu}^2}{32\pi^2},
\la{TA}\eeq
where $\beta(g^2)$ is the Gell-Mann--Low function,

\beq
\beta(g^2) \equiv \frac{dg^2(M)}{\ln M}=-b\frac{g^4}{8\pi^2}-
\frac{b^\prime}{2}\frac{g^6}{(8\pi^2)^2}-...,\;\;\;b=\frac{11}{3}N_c,\;\;
b^\prime=\frac{34}{3}N_c^2,
\la{GML}\eeq
one gets \cite{DP1}:

\beq
\frac{{\cal Z}}{{\cal Z}_{P.T.}}=\exp\left( \frac{b}{4}V
\langle F_{\mu\nu}^2/32\pi^2\rangle_{NP}\right)
\la{GCdef}\eeq
where $\langle F_{\mu\nu}^2\rangle_{NP}$ is the gluon field vacuum
expectation value which is due to non-perturbative fluctuations, i.e.
the gluon condensate. The aim of any QCD-vacuum builder
is to minimize the vacuum energy or, equivalently, to maximize the
gluon condensate.

It is important that it is a
renormalization-invariant quantity \footnote{To be more precise, the
renorm-invariant quantity is $\langle\theta_{\mu\mu}\rangle$,
see \eq{TA}; however if the coupling constant at the ultra-violet cutoff
scale $g^2(M)$ is small enough, it is sufficient to use the beta
function from one loop.}, meaning that its dependence on the ultraviolet
cutoff $M$ and the bare charge $g^2(M)$ given at this cutoff is such
that it is actually cutoff-independent:

\beq
\langle F_{\mu\nu}^2/32\pi^2\rangle_{NP}
=c\left[M\exp\left(-\int_{...}^{g^2(M)}
\frac{dg^2}{\beta(g^2)}\right)\right]^4
\simeq c^\prime M^4\exp\left[-\frac{32\pi^2}{bg^2(M)}\right].
\la{RI}\eeq
By definition of the QCD partition function \ur{partfu},
the l.h.s. of \eq{RI} is equal to $-\frac{1}{V}d\ln
{\cal Z}/d[8\pi^2/g^2(M)]$.  Applying the same differentiation
operator to \eq{RI} one gets a low-energy theorem \cite{NSVZ}:

\[
\frac{d^2\ln \left({\cal Z}/{\cal Z}_{P.T.}\right)}
{\left(d\left[-\frac{8\pi^2}{g^2(M)}\right]\right)^2} =\left\langle
\int d^4x\frac{F_{\mu\nu}^2}{32\pi^2} \int
d^4y\frac{F_{\mu\nu}^2}{32\pi^2}\right\rangle
-\left\langle \int d^4x\frac{F_{\mu\nu}^2}{32\pi^2}\right\rangle^2
\]
\beq
= \frac{4}{b}\left\langle
\int d^4x\frac{F_{\mu\nu}^2}{32\pi^2}\right\rangle.
\la{LET}\eeq
If the bare coupling $g^2(M)$ is not chosen small enough, there are
obvious corrections to this formula, following from the higher-order
terms in the beta function.

This low-energy theorem has an instructive consequence for instantons,
predicting the dispersion of the number of pseudoparticles in a
given 4-dimensional volume \cite{DP1}. Assuming the instanton ensemble
is sufficiently dilute (corrections to this assumption will be
discussed below) one can rewrite the low-energy theorem \ur{LET} as

\beq
\langle N^2\rangle - \langle N\rangle^2
=\frac{4}{b}\langle N\rangle =\frac{12}{11N_c}\langle N\rangle,
\la{NN}\eeq
where $N\equiv N_+ +N_-$ is the total number of \IIs.

Thus it follows from the renormalization properties of the Yang--Mills
theory that the dispersion of the number of pseudoparticles is {\em
less} than for a free gas for which one would get a Poisson
distribution with $\langle N^2\rangle - \langle N\rangle^2 =
\langle N\rangle$. In particular, at $N_c\rightarrow \infty$ the
dispersion becomes zero, as it should be.

One concludes that some kind of interaction of instantons with each
other is crucial to support the needed renormalization properties of
the underlying theory: any cutoff of the integrals over instanton sizes
"by itself" leads to the Poisson distrubution and hence to the
violation of the low energy theorem \ur{NN}.

Differentiating $ln\; {\cal Z}$ many times in respect to the bare
coupling $g^2(M)$, one can easily generalize \eq{NN} to any moments of
the distribution. In short, the distrubution in the number of
pseudoparticles should be (for large $\langle N \rangle $)

\beq
P(N) \sim \exp\left[-\frac{b}{4}\left(\ln\frac{N}{\langle N\rangle}
-1\right)\right].
\la{DN}\eeq

\section{One-instanton weight}
\setcounter{equation}{0}
\def\theequation{\arabic{section}.\arabic{equation}}

The words "instanton vacuum" mean that one assumes that the QCD
partition function is mainly saturated by an ensemble of interacting \IIs
together with quantum fluctuations about them. Instantons are
necessarily present in the QCD vacuum if only because they lower the
vacuum energy in respect to the purely perturbative (divergent) one.
The question is whether they give the dominant contribution to the
gluon condensate, and to other basic quantities. To answer this
question one has to compute the partition function \ur{partfu} assuming
that it is mainly saturated by instantons, and to compare the
obtained gluon condensate with the phenomenological one. This work
has been done a decade ago in ref. \cite{DP1}; today direct lattice
measurements confirm that the answer to the question is positive: the
observed density of \IIs is in agreement with the estimate \ur{avsep}.

The starting point of this calculation is the contribution of one
isolated instanton to the partition function \ur{partfu} or the
one-instanton weight. It has been computed by 't Hooft \cite{tH},
and generalized to arbitrary groups by Bernard in a very clearly
written paper \cite{B}.

The general field can be decomposed as a sum of a classical field of
an instanton $A_\mu^I(x,\xi)$ where $\xi$ is a set of $4N_c$
collective coordinates characterizing a given instanton (see
\eq{instgen}), and of a presumably small quantum field $a_\mu(x)$:

\beq
A_\mu(x)=A_\mu^I(x,\xi) + a_\mu(x).
\la{genfi}\eeq
There is a subtlety in this decomposition due to the gauge freedom:
an interested reader is addressed to ref. \cite{DP1} where this
subtlety is treated in detail.

The action is

\[
{\rm Action}=\frac{1}{4g^2}\int d^4x\; F_{\mu\nu}^2=\frac{8\pi^2}{g^2}
+\frac{1}{g^2}\int d^4x\; D_\mu F_{\mu\nu}a_\nu
\]
\beq
+\frac{1}{2g^2}\int d^4x\; a_\mu W_{\mu\nu}a_\nu + O(a^3).
\la{quaform}\eeq

Here the term linear in $a_\mu$ drops out because the instanton field
satisfies the equation of motion. The quadratic form $W_{\mu\nu}$ has
$4N_c$ zero modes related to the fact that the action does not depend
on $4N_c$ collective coordinates. This brings in a divergence in the
functional integral over the quantum field $a_\mu$ which, however, can
and should be qualified as integrals over the collective coordinates:
centre, size and orientations. Formally the functional integral over
$a_\mu$ gives

\beq
\frac{1}{\sqrt{\det\;W_{\mu\nu}(A^I)}},
\la{funcdet}\eeq
which must be {\it i}) normalized (to the determinant of the free
quadratic form, i.e. with no background field), {\it ii}) regularized
(for example by using the Pauli--Villars method), and {\it iii})
accounted for the zero modes. Actually one has to compute a
"quadrupole" combination,

\beq
\left[\frac{\det^\prime W \; \det (W_0+M^2)}{\det W_0 \; \det
(W+M^2)}\right]^{-\frac{1}{2}},
\la{quadrup}\eeq
where $W_0$ is the quadratic form with no background field and $M^2$ is
the Pauli--Villars mass playing the role of the ultra-violet cutoff;
the prime reminds that the zero modes should be removed and treated
separately. The resulting one-instanton contribution to the partition
function (normalized to the free one) is \cite{tH,B}:

\[
\frac{{\cal Z}_{1-inst}}{{\cal Z}_{P.T.}}
=\int d^4z_\mu\int d\rho\int dU\; d_0(\rho),
\]
\beq
d_0(\rho)=\frac{C(N_c)}{\rho^5}\left[\frac{8\pi^2}{g^2(M)}\right]^{2N_c}
(M\rho)^{\frac{11}{3}N_c}\exp\left(-\frac{8\pi^2}{g^2(M)}\right).
\la{1instw}\eeq

The product of the last two factors is actually a combination of the
cutoff $M$ and the bare coupling constant $g^2(M)$ given at this
cutoff, which is cutoff-independent; it can be replaced by
$(\Lambda_{QCD}\rho)^{11N_c/3}$, see \eq{RI}. This is the way
$\Lambda_{QCD}$ enters into the game; henceforth all dimensional
quantities will be expressed through $\Lambda_{QCD}$, which is, of
course, a welcome message. The numerical coefficient $C(N_c)$ depends
explicitly on the number of colours; it also implicitly depends on the
regularization scheme used. In the Pauli--Villars scheme exploited
above \cite{B}

\beq
C(N_c)=\frac{4.60\exp(-1.68N_c)}{\pi^2(N_c-1)!(N_c-2)!}.
\la{CNc}\eeq
If the scheme is changed, one has to change the coefficient
$C(N_c) \rightarrow C^\prime(N_c) =
C(N_c)\cdot(\Lambda/\Lambda^\prime)^b$. One has: $\Lambda_{P.V.} =
1.09\Lambda_{\bar M\bar S} = 31.32\Lambda_{lat} = ...$

We have thus obtained the instanton weight in the one-loop
approximation:

\beq
d_0(\rho)= \frac{C(N_c)}{\rho^5}\beta(M)^{2N_c}
exp\left[-\beta(\rho)\right],
\la{d01}\eeq
where $\beta(\rho)=8\pi^2/g^2(\rho)$ is the one-loop inverse charge
(not to be confused with the Gell-Mann--Low function!)

\beq
\beta(\rho)= b\ln\left(\frac{1}{\Lambda\rho}\right),\;\;\;\;
b=\frac{11}{3}N_c.
\la{1l}\eeq
Note that the $\beta$ factor in the pre-exponent starts to "run" only
at the 2-loop level, hence its argument is taken at the ultra-violet
cutoff $M$.

In the 2-loop approximation the instanton weight is given by
\cite{VZNS}

\beq
d_0(\rho)= \frac{C(N_c)}{\rho^5}\beta(\rho)^{2N_c}
\exp\left[-\beta^{II}(\rho)+
\left(2N_c-\frac{b^\prime}{2b}\right)\frac{b^\prime}{2b}
\frac{\ln\beta(\rho)}{\beta(\rho)}+O(1/\beta(\rho))
\right],
\la{d02}\eeq
where $\beta^{II}(\rho)$ is the inverse charge to the two-loop
accuracy:

\beq
\beta^{II}(\rho)=\beta(\rho)+\frac{b^\prime}{2b}\ln\frac{2\beta(\rho)}{b},
\;\;\;\;b^\prime=\frac{34}{3}N_c^2.
\la{2l}\eeq

Notice that both one- and two-loop \eqs{d01}{d02} formulae show that
the integral over the instanton sizes $\rho$ in \eq{1instw} diverges
as a high power of $\rho$ at large $\rho$: this is of course the
consequence of asymptotic freedom. It means that individual instantons
tend to swell. This circumstance plagued the instanton calculus for
many years. If one attemts to cut the $\rho$ integrals "by hand", one
violates the renormalization properties of the YM theory, as explained
in the previous section. Actually the size integrals appear to be cut
from above due to instanton interactions.

\section{Instanton ensemble}
\setcounter{equation}{0}
\def\theequation{\arabic{section}.\arabic{equation}}

To get a volume effect from instantons one needs to consider an
\II ensemble, with their total number $N$ proportional to the
4-dimensional volume $V$. Immediately a mathematical difficulty arises:
any superposition of \IIs is not, strictly speaking, a solution of the
equation of motion, therefore, one cannot directly use the
semiclassical approach of the previous section. There are two ways two
overcome this difficulty. One is to use a variational principle
\cite{DP1}, the other is to use the effective YM lagrangian in the
instanton field \cite{DPol}.

The idea of the variational principle is to use a modified YM action
for which a chosen \II ansatz {\em is} a saddle point. Exploiting the
convexity of the exponent one can prove that the true vacuum energy is
{\em less} than that obtained from the modified action. One can
therefore use variational parameters (or even functions) to get a best
upper bound for the vacuum energy. We call it the Feynman variational
principle since the method was suggested by Feynman in his famous
study of the polaron problem. The gauge theory is more difficult,
though: one has not to loose either gauge invariance or the
renormalization properties of the YM theory. These difficulties were
overcome in ref. \cite{DP1}. A decade later I still do not think one
can do an analytical evaluation of the \II ensemble much better than
in that paper: after all we are dealing with the "strong interactions",
meaning that all dimenionless quantities are generally speaking of
the order of unity -- there are no small parameters in the theory.
Therefore, one has to use certain numerical methods, and the
variational principle is among the best. Todays direct lattice
investigation of the \II ensemble seem to indicate that Petrov and I
have obtained rather accurate numbers in this terrible problem.

The normalized (to perturbative) and regularized YM partition
function takes the form of a partition function for a grand canonical
ensemble of interacting psuedoparticles of two kind, \IIs:

\beq
\frac{{\cal Z}}{{\cal Z}_{P.T.}}
= \sum_{N_+,N_-}\frac{1}{N_+!} \frac{1}{N_-!}
\prod_n^{N_+ + N_-} \int d^4 z_n d\rho_n dO_I\; d_0(\rho_n)\;
\exp(-U_{int}),
\la{GPF}\eeq
where $d_0(\rho)$ is the 1-instanton weight \ur{d01} or \ur{d02}.
The integrals are over the collective coordinates of (anti)instantons:
their coordinates $z$, sizes $\rho$ and orientations given by $SU(N_c)$
unitary matrices in the adjoint representation $O$; $dO$ means the
Haar measure normalized to unity. The instanton interaction
potential $U_{int}$ (to be discussed below) depends on the
separation between pseudoparticles, $z_m-z_n$, their sizes $\rho_{m,n}$
and their relative orientations $O_mO_n^T$.

In the variational approach the interaction between instantons arise
from {\em i}) the defect of the classical action, {\em ii}) the
non-factorization of quantum determinants and {\em iii}) the
non-factorization of jacobians when one passes to integartion over the
collective coordinates. All three factors are ansatz-dependent, but
there is a tendency towards a cancellation of the ansatz-dependent
pieces. Qualitatively, in any ansatz the interactions between \IIs
resemble those of molecules: at large separations there is an
attraction, at smaller separations there is a repulsion. It is very
important that the interactions depend on the relative orientations of
instantons: if one averages over orientations (which is the natural
thing to do if the \II medium is in a disordered phase; if not, one
would expect a spontaneous breaking of both Lorentz and colour
symmetries \cite{DP1}), the interactions seem to be repulsive at any
separations.

In general, the mere notion of the instanton interactions is notorious
for being ill-defined since instanton + antiinstanton is not a solution
of the equation of motion. Such a configuration belongs to a sector
with topological charge zero, thus it seems to be impossible to
distinguish it from what is encountered in perturbation theory. The
variational approach uses brute force in dealing with the problem, and
the results appear to be somewhat dependent on the ansatz used. Thanks
to the inequality for the vacuum energy mentioned above, we still get
quite a useful information. However, recently a mathematically
unequivocal definition of the instanton interaction potential has been
suggested, based on analyticity and unitarity \cite{DP4,DPol}. This
definition automatically cuts off the contribution of the perturbation
theory. The first three leading terms for the interaction potential
at large separations has been computed \cite{DPol}, at smaller
separations one observes a strong repulsion \cite{DP5}, though the
exact form is still unknown.

Summing up the discussion, I would say that today there exists no
evidence that a variational calculation with the simplest sum ansatz
used in ref. \cite{DP1} is qualitatively of even quantitatively
incorrect, therefore I will cite the numerics from those cacluations in
what follows.

The main finding is that the \II ensemble stabilizes at a certain
density related to the $\Lambda_{QCD}$ parameter (there is no other
dimensional quantity in the theory!):

\beq
\langle F_{\mu\nu}^2/32\pi^2\rangle\simeq \frac{1}{V}\langle Q_T^2
\rangle\simeq \frac{N}{V}\geq (0.75 \Lambda_{\bar M\bar S})^4
\la{numval}\eeq
which would require $\Lambda_{\bar M\bar S}\simeq 265\;MeV$ to get the
phenomenological value of the condensate. It should be mentioned
however that using more sophisticated variational Ans\"{a}tze one can
obtain a larger coefficient in \eq{numval} and hence would need smaller
values of $\Lambda$.

The average sizes $\bar \rho$ appear to be much less
than the average separation $\bar R$.  Numerically we have found for
the $SU(3)$ colour:

\beq
\frac{\bar \rho}{\bar R} \simeq \frac{1}{3}
\la{pf}\eeq
which coincides with what was suggested previously by Shuryak
\cite{Sh1} from considering the phenomenological applications of the
instanton vacuum. This value should be compared with that found from
direct lattice measurements \cite{CGHN}: $\bar \rho /\bar R \simeq
.37 - .4$, depending on where one stops the cooling procedure. The
packing fraction, i.e. the fraction of the 4-dimensional volume
occupied by instantons apears thus to be rather small, $\pi^2 \bar\rho^4
/\bar R^4 \simeq 1/8$. This small number can be traced back to the
"accidentally" large numbers appearing in the 4-dimensional YM theory:
the $11N_c/3$ of the Gell-Mann--Low beta function and the number of
zero modes being $4N_c$. We have checked that the same variational
principle applied to the 2-dimensional sigma models also possessing
instantons, does not yield a small packing fraction. The 4-dimensional
YM theory seems to be simpler from this angle. Meanwhile, it
is exactly this small packing fraction of the instanton vacuum which
gives an {\it a posteriori} justification for the use of the
semi-classical methods. As I shall show in the next sections, it also
enables one to identify adequate degrees of freedom to describe the
low-energy QCD.

\section{Chiral symmetry breaking in QCD}
\setcounter{equation}{0}
\def\theequation{\arabic{section}.\arabic{equation}}

The QCD lagrangian with $N_f$ massless flavours is known to posses
a large global symmetry, namely a symmetry under $U(N_f)\times U(N_f)$
independent rotations of left- and right-handed quark fields. This
symmetry is called {\em chiral}. Instead of rotating separately the
2-component Weyl spinors one can make independent vector and axial
$U(N_f)$ rotations of the full 4-component Dirac spinors -- the QCD
lagrangian is invariant under these transformations too. Meanwhile, the
axial transformation mixes states with different P-parities. Therefore,
if that symmetry remains exact one would observe parity degeneracy
of all states with otherwise the same quantum numbers. In reality the
splittings between states with the same quantum numbers but opposite
parities is huge. For example, the splitting between the vector $\rho$
and the axial $a_1$ meson is $(1200 - 770)\simeq 400\;MeV$; the
splitting between the nucleon and its parity partner is even larger:
$(1535 - 940)\simeq 600\; MeV$. (Another possibility is that the
nucleon is just massless, which looks even less pleasant.)

The splittings are too large to be explained by the small bare or
current quark masses which break the chiral symmetry from the
beginning. Indeed, the current masses of light quarks are: $m_u \simeq
4\;MeV,\;\;m_d\simeq 7\;MeV,\;\;m_s\simeq 150\;MeV$. The only
conclusion one can draw from these numbers is that the chiral symmetry
of the QCD lagrangian is broken down {\em spontaneously}, and very
strongly. Consequently, one should have light (pseudo) Goldstone
pseudoscalar hadrons -- their role is played by pions which indeed are
by far the lightest hadrons.

The order parameter associated with chiral symmetry breaking is
the so-called {\em chiral} or {\em quark condensate}:

\beq
\langle\bar\psi\psi\rangle\simeq -(250\;MeV)^3.
\la{chcond}\eeq
It should be noted that this quantity is well defined only for
massless quarks, otherwise it is somewhat ambigious. By definition,
this is the quark Green function taken at one point; in momentum space
it is a closed quark loop. If the quark propagator has only the "slash"
term, the trace over the spinor indices implied in this loop would give
an identical zero. Therefore, chiral symmetry breaking implies that
a massless (or nearly massless) quark develops a non-zero dynamical
mass (i.e. a "non-slash" term in the propagator). There are no reasons
for this quantity to be a constant independent of the momentum;
moreover, we understand that it should anyhow vanish at large momentum.
The value of the dynamical mass at zero momentum can be estimated as
one half of the $\rho$ meson mass or one third of the nucleon mass,
that is about $M(0)\simeq 350-400\;MeV$; this scale is also related
to chiral symmetry breaking and should be emerge together with the
condensate \ur{chcond}.

One could imagine a world without confinement but with chiral symmetry
breaking: it would not be drastically different from what we meet in
reality. There would be a tightly bound light Goldstone pion, and
relatively loosely bound $\rho$ meson and nucleon with correct masses,
which, however, would be possible to ionize from time to time. Probably
the spectrum of the highly excited hadrons would be wrong, though even
that is not so clear \cite{DP6}.
We see, thus, that the spontaneous chiral symmetry breaking is the main
dynamical happening in QCD, which determines the face of the strong
interactions world. In the next sections I explain why and how chiral
symmetry is broken in the instanton vacuum, and why it is a most
realistic picture. The forthcoming sections are based on our work with
Petrov \cite{DP2,DP3}.

\section{Chiral symmetry breaking by definition}
\setcounter{equation}{0}
\def\theequation{\arabic{section}.\arabic{equation}}

I start by writing down the QCD partition function. Functional
integrals are well defined in euclidean space which is obtained by the
following formal substitutions of Minkowski space quantitites:

\[
ix_{M0}=x_{E4},\;\;\;x_{Mi}=x_{Ei},\;\;\;A_{M0}=iA_{E4},\;\;\;A_{Mi}=A_{Ei},
\]
\beq
i\bar\psi_M=\psi_E^\dagger,\;\;\;\gamma_{M0}=\gamma_{E4},\;\;\;
\gamma_{Mi}=i\gamma_{Ei},\;\;\;\gamma_{M5}=\gamma_{E5}.
\la{ME}\eeq

Neglecting for brevity the gauge fixing and Faddeev--Popov ghost terms,
the QCD partition function can be written as

\[
{\cal Z}=\int DA_\mu D\psi D\psi^\dagger\;
\exp\left[-\frac{1}{4g^2}\int F_{\mu\nu}^2+\sum_f^{N_f}
\int\psi_f^\dagger(i\nabla+im_f)\psi_f\right]
\]
\beq
= \int DA_\mu \exp\left[-\frac{1}{4g^2}\int F_{\mu\nu}^2\right]
\prod_f^{N_f}\det(i\nabla + im_f).
\la{partfuq}\eeq

The chiral condensate of a given flavour $f$ is, by definition,

\beq
\langle\bar\psi_f\psi_f\rangle_M=-i\langle\psi_f^\dagger\psi_f\rangle_E
=-\frac{1}{V}\frac{\partial}{\partial m_f}
\left(\ln {\cal Z}\right)_{m_f\rightarrow 0}.
\la{ccdef}\eeq

The Dirac operator has the form

\beq
i\nabla=\gamma_\mu(i\partial_\mu+A_\mu^{I\bar I}+a_\mu)
\la{Dirop}\eeq
where $A_\mu^{I\bar I}$ denotes the classical field of the \II
ensemble and $a_\mu$ is a presumably small field of quantum
fluctuations about that ensemble, which I shall neglect as it has
little impact on chiral symmetry breaking. Integrating over $DA_\mu$
in \eq{partfuq} means averaging over the \II ensemble, therefore one
can write

\beq
{\cal Z} = \overline{\det(i\nabla+im)}
\la{zav}\eeq
where I temporarily restrict the discussion to the case of only one
flavour for simplicity. Because of the $im$ term the Dirac operator
in \ur{zav} is formally not hermitean; however the determinant is real
due to the following observation. Suppose we have found the eigenvalues
and eigenfunctions of the Dirac operator,

\beq
i\nabla\Phi_n=\lambda_n\Phi_n,
\la{eig}\eeq
then for any $\lambda_n \neq 0$ there is an eigenfunction
$\Phi_{n^\prime} =\gamma_5\Phi_n$ whose eigenvalue is
$\lambda_{n^\prime}=-\lambda_n$.  This is because $\gamma_5$
anticommutes with $i\nabla$. Owing to this the fermion determinant can
be written as

\[
\det(i\nabla+im)=\prod_n(\lambda_n+im)=\sqrt{\prod(\lambda_n^2+m^2)}
=\exp\left[\frac{1}{2}\sum_n\ln(\lambda_n^2+m^2)\right]
\]
\beq
=\exp\left[\frac{1}{2}\int_{-\infty}^\infty d\lambda\;\nu(\lambda)
\ln(\lambda^2+m^2)\right],
\;\;\;\nu(\lambda)\equiv \sum_n\delta(\lambda-\lambda_n),
\la{fermdet}\eeq
where I have introduced the {\em spectral density} $\nu(\lambda)$ of
the Dirac operator $i\nabla$. Note that the last expression is real and
even in $m$, which is a manifestation of the QCD chiral invariance.

Differentiating \eq{fermdet} in $m$ and putting it to zero one gets
according to the general \eq{ccdef} a formula for the chiral
condensate:

\beq
\langle\bar\psi\psi\rangle= -\frac{1}{V}\frac{\partial}{\partial m}
\left[\frac{1}{2}\int d\lambda\;
\overline{\nu(\lambda)}\ln(\lambda^2+m^2)\right]_{m\rightarrow 0}
=\left.-\frac{1}{V}\int_{-\infty}^\infty d\lambda\;
\overline{\nu(\lambda)}\frac{m}{\lambda^2+m^2)}\right|_{m\rightarrow 0}
\la{BC1}\eeq
where $\overline{\nu(\lambda)}$ means averaging over the instanton
ensemble together with the weight given by the fermion determinant
itself. The latter, however, may be cancelled in the so-called
quenched approximation where the back influence of quarks on the
dynamics is neglected. Theoretically, this is justified at large $N_c$.

Naively, one would think that the r.h.s. of \eq{BC1} iz zero at
$m\rightarrow 0$. That would be correct for a finite-volume system with
a discrete spectral density. However, if the volume goes to infinity
faster than $m$ goes to zero (which is what one should assume in the
thermodynamic limit) one should use instead

\beq
\frac{m}{\lambda^2+m^2}\stackrel{m\rightarrow 0}{\longrightarrow}
{\rm sign}(m)\pi\delta(\lambda),
\la{thl}\eeq
so that one gets \cite{BC}:

\beq
\langle\bar\psi\psi\rangle=-\frac{1}{V}{\rm sign}(m)
\pi\overline{\nu(0)}.
\la{BC2}\eeq

The chiral condensate is thus proportional to the averaged spectral
density of the Dirac operator at zero eigenvalues. The appearance of
the sign function is not accidental: it means that at small $m$ QCD
partition functions depends on $m$ non-analytically:

\beq
\ln{\cal Z}= V(c_0+\pi\overline{\nu(0)}|m|+c_2m^2\ln(|m|)+...).
\la{nonan}\eeq

The fact that the partition function is even in $m$ is the reflection
of the original invariance of the QCD under $\gamma_5$ rotations; the
fact that it is non-analytic in the symmetry-breaking parameter $m$
is typical in the situation where symmetry is broken spontaneously.

A generalization of the above formulae to the case of several flavours
is simple \cite{D2}. Consider $N_f$ quark flavours with a most general
mass matrix

\beq
\psi^\dagger\left(im_L\frac{1+\gamma_5}{2}+im_R\frac{1-\gamma_5}{2}
\right)\psi.
\la{genmass}\eeq
The fermion determinant can be written as

\[
\det(i\nabla+im)
=\exp\left[\frac{1}{2}\int d\lambda \nu(\lambda)
\ln {\rm det}_{N_f}(\lambda^2+m_Lm_R)\right]
\]
\beq
=\exp\left[-\frac{1}{2}\int d\lambda\; \nu(\lambda)\int\frac{dt}{t}
{\rm Tr}_{N_f}\exp\left(-t(\lambda^2+m_Lm_R)\right)\right]
\la{ferdetge}\eeq
where $\lambda^2$ is a $N_f\times N_f$ matrix proportional to the unit
matrix. Let us expand $\nu(\lambda)$ at small $\lambda$, and leave only
the $\nu(0)$ term. Integrating \ur{ferdetge} in $\lambda$ we get the
correspondent non-analytic term in the partition function:

\beq
\ln{\cal Z} \sim
-\frac{\sqrt{\pi}}{2}\nu(0)\int_0^\infty\frac{dt}{t^{3/2}}{\rm Tr}_{N_f}
\left[\exp(-tm_Lm_R)\right]=\pi\nu(0){\rm Tr}_{N_f}(m_Lm_R)^{1/2}.
\la{nonange}\eeq
This expression is non-analytic in the mass matrix; differentiating it
in respect to the masses one finds the phases of the condensate.

In the next sections I shall present three different derivations of the
fact that instantons indeed break chiral symmetry.

\section{Chiral symmetry breaking by instantons: qualitative
derivation}
\setcounter{equation}{0}
\def\theequation{\arabic{section}.\arabic{equation}}

The key observation is that the Dirac operator in the background field
of one (anti) instanton has an exact zero mode with $\lambda=0$
\cite{tH}. It is a consequence of the general Atiah--Singer
index theorem; in our case it is guaranteed by the unit Pontryagin
index or the topological charge of the instanton field. These
zero modes are 2-component Weyl spinors: {\em right}-handed for
instantons and {\em left}-handed for antiinstantons. Explicitly, the
zero modes are ($\alpha=1...N_c$ is the colour and $i,j,k=1,2$ are the
spinor indices):

\[
\left[\Phi_R(x-z_1)\right]^\alpha_i=\phi(x-z_1,\rho_1)(x-z_1)^+_{ij}
U_{1k}^\alpha\epsilon^{jk},
\]
\[
\left[\Phi_L(x-z_2)\right]^\alpha_i=\phi(x-z_2,\rho_2)(x-z_2)^-_{ij}
U_{2k}^\alpha\epsilon^{jk},
\]
\beq
\phi(x,\rho)=\frac{\rho}{\pi(2x^2)^{1/2}(x^2+\rho^2)^{3/2}}.
\la{zm}\eeq
Here $z_{1\mu},\rho_1,U_1$ are the center, size and orientation of
an instanton and $z_{2\mu},\rho_2,U_2$ are those of an antiinstanton,
respectively, $\epsilon^{jk}$ is the $2\times 2$ antisymmetric matrix.

For infinitely separated $I$ and $\bar I$ one has thus
two degenerate states with exactly zero eigenvalues. As usual in
quantum mechanics, this degeneracy is lifted through the
diagonalization of the hamiltonian, in this case the hamiltonian is
the full Dirac operator. The two "wave functions" which diagonalize the
"hamiltonian" are the sum and the difference of the would-be zero
modes, one of which is a 2-component left-handed spinor, and the other
is a 2-component right-handed spinor. The resulting wave functions are
4-component Dirac spinors; one can be obtained fron another by
multiplying by the $\gamma_5$ matrix.  As the result the two would-be
zero eigenstates are split symmetrically into two $4$-component Dirac
states with {\em non-zero} eigenvalues equal to the overlap integral
between the original states:

\beq
\lambda=\pm |T_{12}|,\;\;\;\;
T_{12}=\int d^4x \Phi_1^\dagger(-i\!\not\partial)\Phi_2
\stackrel{R_{12}\rightarrow\infty}{\longrightarrow}
-\frac{2\rho_1\rho_2}{R_{12}^4}{\mbox Tr}(U_1^\dagger U_2R_{12}^+).
\la{overl}\eeq

We see that the splitting between the would-be zero modes fall off as
the third power of the distance between $I$ and $\bar I$; it also
depends on their relative orientation. The fact that two levels have
eigenvalues $\pm\lambda$ is in perfect agreement with the $\gamma_5$
invariance mentioned in the previous section.

When one adds more \IIs each of them brings in a would-be zero
mode. After the diagonalization they get split symmetrically in
respect to the $\lambda=0$ axis.  Eventually, for an \II ensemble one
gets a continuous band spectrum with a spectral density $\nu(\lambda)$
which is even in $\lambda$ and finite at $\lambda=0$.

One can make a quick estimate of $\nu(0)$: Let the total number
of \IIs in the 4-dimensional volume $V$ be $N$. The spread $\Delta$ of
the band spectrum of the would-be zero modes is given by their average
overlap \ur{overl}:

\beq
\Delta\sim\sqrt{\int (d^4R/V) T(R)T^*(R)}\sim\frac{\bar\rho}{\bar R^2}
\la{averoverl}\eeq
where $\bar\rho$ is the average size and $\bar R=(N/V)^{-1/4}$ is the
average separation of instantons. Note that the spread of the
would-be zero modes is parametrically much less than $1/\bar\rho$ which
is the typical scale for the non-zero modes. Therefore, neglecting the
influence of the non-zero modes is justified if the packing
fraction of instantons is small enough. From \eq{BC2} one gets an
estimate for the chiral condensate induced by instantons:

\beq
\langle\bar\psi\psi\rangle=-\frac{\pi}{V}\nu(0)
\simeq -\frac{\pi}{V}\frac{N}{\Delta}\sim-\frac{1}{\bar R^2 \bar\rho}.
\la{ccest}\eeq
Note that the chiral condensate appears to be proportional to the {\em
square root} of the instanton density $N/V$: again it is as it should
be for the order parameter of spontaneous symmetry breaking.

It is amusing that the physics of the spontaneous breaking of chiral
symmetry resembles the so-called Mott--Anderson conductivity in
disordered systems. Imagine random impurities (atoms) spread over a
sample with final density, such that each atom has a localized bound
state for an electron. Due to the overlap of those localized electron
states belonging to individual atoms, the levels are split into a band,
and the electrons become delocalized. That means conductivity of the
sample. In our case the localized zero quark modes of individual
instantons randomly spread over the volume get delocalized due to their
overlap, which means chiral symmetry breaking.

There is a difference with the Mott--Anderson conductivity, though.
In the case of atoms, the wave functions have an exponential falloff,
so that the overlap integrals are exponentially small at large
separations. It means that, if the density of impurities is small
enough, there might be a phase transition to an insulator state. In our
case the eigenvalue is exactly zero, and the zero modes decay as a
power -- see \eq{zm}. The overlap integrals are also decreasing as a
power of the separation (see \eq{overl}), and chiral symmetry breaking
occurs at any instanton densities -- at least in the quenched
approximation, when one neglects the back influence of quarks on the
dynamics of instantons. However, at non-zero temperatures the density
of instantons decrease \cite{DM}, and the fermion zero modes have an
exponential falloff \cite{KY}; therefore one can expect a
Mott--Anderson phase transition to an insulator state, meaning the
restoration of chiral symmetry. It should be noted that chiral symmetry
may be restored as due to the back influence of quarks on the instanton
ensemble -- it is an $O(N_f/N_c)$ effect then. It has been studied
recently in refs. \cite {Sh4} where a formation of instanton molecules
has been suggested as a mechanism of chiral symmetry restoration at
high temperatures. In fact, the restoration mechanism may be different
for different numbers of colours and flavours.

I should mention that the idea that instantons can break chiral
symmetry has been discussed previously (see  refs. \cite{C,CC,CDG,Sh1})
however the present mechanism and a consistent formalism has been
suggested and developed in papers \cite{DP2,DP3}.

Recently there have been much interesting work done generalizing this
mechanism of chiral symmetry breaking to other configurations
\cite{GAS} and studying general properties of the spectral density
of the Dirac operator for various ensembles \cite{LS,V}.

\section{Derivation II: quark propagator}
\setcounter{equation}{0}
\def\theequation{\arabic{section}.\arabic{equation}}

Having explained the physical mechanism of chiral symmetry breaking as
due to the delocalization of the would-be zero fermion modes in the
field of individual instantons, I shall indicate how to compute
observables in the instanton vacuum. The main quantity is the quark
propagator in the instanton vacuum, averaged over the instanton
ensemble. This quantity has been calculated in refs. \cite{DP2,Pob}.
In particular, Pobylitsa \cite{Pob} has derived a closed equation
for the averaged quark propagator, which can be solved as a series
expansion in the formal parameter $N\bar\rho^4/VN_c$ which numerically
is something like 1/250.

The result of refs. \cite{DP2,Pob} is that in the leading order in the
above parameter the quark propagator has the form of a massive
propagator with a momentum-dependent dynamical mass:

\beq
S(p)=\frac{\not p +iM(p^2)}{p^2+M^2(p^2)}, \;\;\;M(p^2)
=c\sqrt{\frac{\pi^2N\bar\rho^2}{VN_c}}F(p\bar\rho),
\la{propag}\eeq
where $F(z)$ is a combination of the modified Bessel functions and
is related to the Fourier transform of the zero mode \ur{zm}: it is
equal to one at $z=0$ and decreases rapidly with the momentum, measured
in units of the inverse average size of instantons (see next
section); $c$ is a constant of the order of unity which depends
slightly on the approximation used in deriving the propagator. Note
that the dynamical quark mass is non-analytical in the instanton
density (similar to the chiral condensate it is an order parameter for
spontaneous symmetry breaking).

Fixing the average  density by the empirical gluon condensate (see
section 4) so that $\bar R\simeq 1\;fm$ and fixing the ratio
$\bar\rho/\bar R = 1/3$ from our variational estimate, we get the value
of the dynamical mass at zero momentum,

\beq
M(0) \simeq 350\;MeV
\la{dynmass}\eeq
while the quark condensate is

\beq
\langle\bar\psi\psi\rangle=i\int\frac{d^4p}{(2\pi)^4}{\rm Tr} S(p)
=-4N_c\int\frac{d^4p}{(2\pi)^4}\frac{M(p)}{p^2+M^2(p)}
\simeq - (255\;MeV)^3.
\la{ccn}\eeq
Both numbers, \ur{dynmass} and \ur{ccn}, appear to be close to their
phenomenological values.

Using the above small parameter one can also compute more complicated
quantities like 2- or 3-point mesonic correlation functions of the type

\beq
\langle J_A(x)J_B(y)\rangle, \;\;\;\langle J_A(x)J_B(y)J_c(z)\rangle,
\;\;\;J_A=\bar\psi\Gamma_A\psi
\la{mescorr}\eeq
where $\Gamma_A$ is a unit matrix in colour but an arbitrary matrix in
flavour and spin. Instantons influence the correlation function in
two ways: {\it i}) the quark and antiquark propagators get dressed and
obtain the dynamical mass, as in \eq{propag}, {\it ii}) quark and
antiquark may scatter simultaneously on the same pseudoparticle;
that leads to certain effective quark interactions. These interactions
are strongly dependent on the quark-antiquark quantum numbers: they are
strong and attractive in the scalar and especially in the pseudoscalar
and the axial channels, and rather weak in the vector and tensor
channels.  I shall derive these interactions in the next section, but
already now we can discuss the pseudoscalar and the axial isovector
channels. These are the channels where the pion shows up as an
intermediate state.

Since we have already obtained chiral symmetry breaking by studying a
single quark propagator in the instanton vacuum, we are doomed to have
a massless Goldstone pion in the appropriate correlation functions.
However, it is instructive to follow how does the Goldstone theorem
manifest itself in the instanton vacuum. It appears that
technologically it follows from a kind of detailed balance in the
pseudoscalar channel (such kind of equations are encountered
in perturbative QCD where there is a delicate cancellation between
real and virtual gluon emission). However, since we have a concrete
dynamical realization of chiral symmetry breaking we can not only
check the general Ward identities of the PCAC (which work of course)
but we are in a position to find quantities whose values do not
follow from general relations. One of the most important quantities is
the $F_\pi$ constant: it can be calculated as the residue of the pion
pole. We get:

\beq
F_\pi=\frac{{\rm const}}{\bar\rho}\left(\frac{\bar\rho}{\bar
R}\right)^2\sqrt{\ln\frac{\bar R}{\bar\rho}}\simeq 100\;MeV\;\;\;{\rm
vs.}\;\; 93\;MeV\;\;({\rm exper.}).
\la{Fpi}\eeq
This is a very instructive formula. The point is, $F_\pi$ is
surprisingly small in the strong interactions scale which, in the
instanton vacuum, is given by the average size of pseudoparticles,
$1/\bar\rho \simeq 600\;MeV$. The above formula says that $F_\pi$ is
down by the packing fraction factor $(\bar\rho/\bar R)^2\simeq 1/9$.
It can be said that $F_\pi$ measures the diluteness of the instanton
vacuum! However it would be wrong to say that instantons are in a
dilute gas phase -- the interactions are crucial to stabilize the
medium and to support the known renormalization properties of the
theory, therefore they are rather in a liquid phase, however dilute it
may turn to be.

By calculating three-point correlation functions in the instanton
vacuum we are able to determine, e.g. the charge radius of the
Goldstone excitation:

\beq
\sqrt{r_\pi^2}\simeq \frac{\surd N_c}{2\pi F_\pi}\simeq (340\;MeV)^{-1}
\;\;\;{\rm vs.}\;\;(310\;MeV)^{-1}\;\;({\rm exper.}).
\la{chra}\eeq

Let me note that all quantities exhibit the natural behaviour in the
number of colours $N_c$:

\[
\langle F_{\mu\nu}^2\rangle \sim \frac{N}{V} = O(N_c),\;\;\;
\langle \bar\psi \psi\rangle = O(N_c),\;\;\;F_\pi^2= O(N_c),
\]
\beq
\bar\rho=O(1),\;\;\;M(0)=O(1),\;\;\;\surd r_\pi^2 =O(1),\;\;
{\rm etc.}
\la{laNc}\eeq

A systematic numerical study of various correlation functions in the
instanton vacuum has been performed by Shuryak, Verbaarschot
and Schaefer \cite{Sh2}, see also Shuryak's lectures at this School.
In all cases considered the results agree well or very well with
experiments and phenomenology. As I already mentioned in the
introduction, similar conclusions have been recently obtained from
direct lattice measurements \cite{CGHN}. I think that one can
conclude that instantons {\em are} explaining the basic properties of
the QCD ground state and that of light hadrons.

\section{Derivation III: Nambu--Jona-Lasinio model}
\setcounter{equation}{0}
\def\theequation{\arabic{section}.\arabic{equation}}

The idea of the first two derivations of chiral symmetry breaking by
instantons, presented above, is: "Calculate quark observables in a
given  background gluon field, then average over the ensemble of
fields", in our case the ensemble of \IIs. The idea of the third
derivation is the opposite: "First average over the \II ensemble
and obtain an effective theory written in terms of interacting quarks
only. Then compute observables from this effective theory". This
approach is in a sense more economical; it has been developed in refs.
\cite{DP3,D2}.

Quark interaction arises when two or more (anti) quarks happen to
scatter over the same pseudoparticle; averaging over its positions and
orientations results in a four- (or more) fermion interaction term
whose range is that of the average size of instantons. The most
essential way how instantons influence quarks is, of course, via the
zero modes. Since each massless quark flavour has its own zero mode,
it means that the effective quark interactions will be actually
$2N_f$ fermion ones. They are usually referred to as {\em 't Hooft
interactions} as he was the first to point out the quantum numbers of
these effective instanton-induced interactions \cite{tH}. In case of
two flavours they are four-fermion interactions, and the resulting
low-energy theory resembles the old Vaks--Larkin--Nambu--Jona-Lasinio
model \cite{VLNJL} which is known to lead to chiral symmetry breaking.
In this section I derive this model from instantons, following refs.
\cite{DP3,D2}. Recently it has been revisited in ref. \cite{DPW}.

The starting point is the quark Green function in the field of one
instanton. It can be written as a sum over all eigenfunctions of the
correspondent Dirac operator $\Phi_n(x)$ -- see \eq{eig}:

\[
S^I(x,y)\equiv \langle \psi(x)\psi^\dagger(y)\rangle
= -\sum_n\frac{\Phi_n(x)\Phi_n^\dagger(y)}{\lambda_n+im}
\]
\beq
=-\frac{\Phi_0(x)\Phi_0^\dagger(y)}{im}+S^{\prime I}(x,y)
\la{qGf}\eeq
where $\Phi_0$ is the zero mode \ur{zm} (henceforth I omit the
subscript $0$), and $S^\prime$ is the sum over non-zero modes,
which is finite in the chiral limit $m\rightarrow 0$. For the
simplicity of the derivation we replace it by the free Green
function $S_0(x,y)$, though the exact propagator in the field of one
instanton is known. Thus instead of the exact propagator we write

\beq
\tilde S^I(x,y)=-\frac{\Phi(x)\Phi^\dagger(y)}{im}+S_0(x,y).
\la{aqGf}\eeq
This approximate Green function is correctly taking into account the
zero mode, that is the low-momentum part, and at large momentum it
reduces to the free Green function, as it should. Therefore, it is
an interpolation of the exact propagator; at momenta $p\sim 1/\rho$
the numerics will be not exact. However, phenomena related to chiral
symmetry breaking correspond to lower momenta, and the use of the
simplification \ur{aqGf} is therefore theoretically justified.

We have now to build the Green function (and other quantities) in the
field of $N_+$ $I$'s and $N_-$ $\bar I$'s and to average over their
ensemble. To that end we use the following mathematical trick.
Consider a fermion action

\[
\exp\left(- A^I[\psi^\dagger, \psi]\right)
=\exp\left(\int d^x \psi^\dagger i\!\not \partial
\psi\right)\left(im-V^I[\psi^\dagger, \psi]\right),
\]
\beq
V^I[\psi^\dagger,\psi]=\int
d^4x\left(\psi^\dagger(x)i\!\not\partial\Phi^I(x) \right)\int
d^4y\left(\Phi^{I\dagger}(y)i\!\not\partial\psi(y)\right)
\la{fermAI}\eeq
where $\Phi^I$ is the zero mode in the field of the $I$th
pseudoparticle. In case of $N_f>1$ one has to take here a product of
the $(im_f-V^I)$ factors for all flavours.

The action \ur{fermAI} has the following properties: {\em i}) the
correspondent partition function normalized to the free one is equal to
$im$, as it should be in case of one instanton; {\em ii}) the Green
function, computed with this action coincides with that of \eq{aqGf}.
It means that this action correctly describes quarks in the field of a
given instanton at low and at large momenta, and interpolates in
between.

In the field of $N_+$ $I$'s and $N_-$ $\bar I$'s the fermion action is

\beq
\exp\left(- A[\psi^\dagger, \psi]\right)
=\prod_f\exp\left(\int d^4x \psi_f^\dagger i\!\not \partial
\psi_f\right)\prod_I^{N_++N_-}\left(im_f-V^I[\psi_f^\dagger,
\psi_f]\right).
\la{fermA}\eeq
To get the QCD partition function one has to integrate over the quark
fields and average over the ensemble of instantons:

\beq
{\cal Z}_{QCD}=\int D\psi D\psi^\dagger
\langle\exp(-A[\psi^\dagger,\psi])\rangle
\la{pfq}\eeq
where $\langle...\rangle$ denotes averaging over the ensemble. It was
shown in ref. \cite{DP1} (see also \cite{DPW}) that the \II ensemble
can be described by an effective {\em one}-particle distribution in the
instanton sizes, which can be found from a variational principle:

\beq
d(\rho)={\rm const}\rho^{b-5}\exp\left(-\frac{\rho^2}{\bar\rho^2}
\frac{b-4}{2}\right),\;\;\;b=\frac{11}{3}N_c-\frac{2}{3}N_f.
\la{sizedi}\eeq
At large $N_c$ it is $\delta$-peaked around the average $\bar\rho$.
Therefore, one can replace the averaging over ensemble by independent
averaging over positions and orientations of invidual pseudoparticles.
Thus \eq{pfq} becomes

\[
{\cal Z}_{QCD}=\int D\psi D\psi^\dagger \exp\left(\int d^4x
\psi_f^\dagger i\!\not\partial\psi_f\right)
\]
\beq
\cdot\prod_f^{N_f}
\left(im_f-\overline{V[\psi_f^\dagger,\psi_f]}\right)^{N_+}
\left(im_f-\overline{V[\psi_f^\dagger,\psi_f]}\right)^{N_-}
\la{pfq1}\eeq
where the bar means averaging over individual pseudoparticles. For
simplicity I shall replace all sizes by their average $\bar\rho$. I
shall also consider only the chiral limit $m_f\rightarrow 0$. It is
natural to introduce non-local $2N_f$ fermion vertices:

\beq
Y_{\pm}=(-)^{N_f}\int d^4z_{I(\bar I)}\int dU_{I(\bar I)}
\prod_f^{N_f}V^{I(\bar I)}[\psi_f^\dagger, \psi_f]
\la{Y}\eeq
where $V^{I(\bar I)}$ depends on the (anti) instanton centers $z_\mu$
and orientations $U$ through the zero modes $\Phi^{I(\bar I)}$, see
\eq{fermAI} and \ur{zm}. The partition function \ur{pfq1} can be
written as

\[
{\cal Z}_{QCD}=\int D\psi_f D\psi_f^\dagger \exp\left(\int d^4x
\psi_f^\dagger i\!\not\partial\psi_f\right)
\]
\beq
\cdot\int \frac{d\lambda_\pm}{2\pi}
\int d\Gamma_\pm \exp\left[i\lambda_+(Y_+-\Gamma_+)
+N_+\ln\frac{\Gamma_+}{V}\; +\; (+\rightarrow -) \right].
\la{pfq2}\eeq
Indeed, integarting over $\lambda_\pm$ one
gets $\delta(Y-\Gamma)$, and after integrating over $\Gamma$ one
recovers \eq{pfq1}. In the thermodynamic limit
$N_\pm,\;V\rightarrow\infty$, $N/V$ fixed, integration over
$\Gamma_\pm$ and $\lambda_\pm$ can be performed by the saddle point
method. We integrate first over $\Gamma_\pm$:

\[
{\cal Z}_{QCD}=\int \frac{d\lambda_\pm}{2\pi}\exp\left[
N_+\left(\ln\frac{N_+}{i\lambda_+V}-1\right)+\;+\;(+\rightarrow
-)\right]
\]
\beq
\cdot\int D\psi_f D\psi_f^\dagger \exp\left(\int d^4x
\psi_f^\dagger i\!\not\partial\psi_f +
i\lambda_+Y_++i\lambda_-Y_-\right).
\la{pfq3}\eeq

Because of the non-locality it is more convenient to write the vertices
\ur{Y} in the momentum space. Let us decompose the 4-component Dirac
spinors describing quark fields into left- and right-handed 2-component
Weyl spinors which we denote as

\beq
\psi_{L(R)}^{f\alpha i},\;\;\; \psi_{L(R)f\alpha i}^\dagger,
\la{We}\eeq
where $f=1...N_f$ stand for flavour, $\alpha=1...N_c$ stand for colour
and $i=1,2$ stand for spin indices. Let us introduce the formfactor
functions $F(k)$ which are related to the Fourier transforms of the
zero modes \ur{zm} and are attributed to each fermion entering the
vertex ($z=k\rho/2$):

\beq
F(k)=2z[I_0(z)K_1(z)-I_1(z)K_0(z)-\frac{1}{z}I_1(z)K_1(z)
\stackrel{k \rightarrow\infty}{\longrightarrow} \frac{6}{k^3\rho^3},
\;\;\;F(0)=1.
\la{Ff}\eeq
The $N_f$ fermion vertices can be written as:

\[
Y_{N_f}^+=\int dU\prod_n^{N_f}\int\frac{d^4k_n}{(2\pi)^4}2\pi\rho F(k_n)
\int\frac{d^4l_n}{(2\pi)^4}2\pi\rho F(l_n)
(2\pi)^4\delta(k_1+...+k_{N_f}-l_1-...-l_{N_f})
\]
\beq
\cdot U_{\gamma_n}^{\alpha_n}U_{\delta_n}^{\dagger\beta_n}
\epsilon^{i_n\gamma_n}\epsilon_{j_n\beta_n}
\left(-\psi_{Lf_n\alpha_ni_n}^\dagger(k_n)\psi_L^{g_n\delta_nj_n}(l_n)
\right);
\la{lY}\eeq
for the vertices $Y^{(-)}$ induced by $\bar I$'s one has to replace
left-handed Weyl spinors by rigth-handed ones. The integral $\int dU$
means averaging over instanton orientations in colour space. In
particlular, one has

\beq
\int dU=1,\;\;\;\int dU\:U_\gamma^\alpha U_\delta^{\dagger\beta}
=\frac{1}{N_c}\delta_\delta^\alpha\delta_\gamma^\beta,\;\;{\rm etc.}
\la{avor}\eeq

To get the $2N_f$ vertices in a closed form one has to perform
explicitly integration over the instanton orientations.  I present
below the results \cite{DP3} for $N_f=1,2$ and for any $N_f$ but
$N_c\rightarrow\infty$.

\vspace{.5cm}
\underline{$N_f=1$}

In this case the "vertex" \ur{Y} is just a mass term:

\beq
Y^\pm_1=\int\frac{d^4k}{(2\pi)^4}\psi^\dagger(k)(2\pi\rho F(k))^2
\frac{1\pm\gamma_5}{2}\psi(k).
\la{Y1}\eeq
One has to plug it into the \eq{pfq3}, integrate over fermions, and
find the saddle-point values of $\lambda_\pm$. If the $CP$ symmetry is
conserved, one has $N_+=N_-=N/2$, and the saddle-point values satisfy
$\lambda_+=\lambda_-$. Then the $\gamma_5$ term in $Y^\pm$ get
cancelled, and \eq{Y1} gives a momentum-dependent mass

\beq
M(k)=M(0)F^2(k),\;\;\;M(0)=\lambda (2\pi\rho)^2,
\la{dmass}\eeq
where $M(0)$ or $\lambda$ is found from the equation
\cite{DP2,DP3} (called sometimes self-consistency or gap equation):

\beq
\frac{4N_c}{N/V}\int\frac{d^4k}{(2\pi)^4}\frac{M^2(k)}{k^2+M^2(k)}=1.
\la{gap}\eeq
Let me mention that exactly the same gap equation \ur{gap}
has been obtained \cite{DP2} in another approach: by first finding the
quark propagator in the instanton vacuum and then averaging over the
instanton ensemble. It is seen from \eq{gap} that the dynamically
generated mass is of the order of $M(0)\sim \sqrt{N/(VN_c)}\bar\rho$.
Knowing the form of $M(k)$ given by \eq{Ff} and using the "standard"
values $N/V=(1\;{\rm fm})^{-4},\; \bar\rho=(1/3)$ fm we find
numerically $M(0)\simeq 350$ MeV. If one neglects $M^2$ in the
denominator of \eq{gap} \cite{Pob} one gets $M(0)\simeq 420$ MeV. This
deviation indicates the accuracy of the "zero mode approximation"
used in this derivation: it is about 15\%.

\vspace{.5cm}
\underline{$N_f=2$}

In this case averaging over the instanton orientations gives a
nontrivial 4-fermion interaction:

\[
Y_2^{(+)}=\frac{2N_c^2}{N/V}\int\frac{d^4k_1d^4k_2d^4l_1d^4l_2}{(2\pi)^{12}}
\sqrt{M(k_1)M(k_2)M(l_1)M(l_2)}
\]
\[
\cdot\frac{\epsilon^{f_1f_2}\epsilon_{g_1g_2}}{2(N_c^2-1)}\left[
\frac{2N_c-1}{2N_c}(\psi_{Lf_1}^\dagger(k_1)\psi_L^{g_1}(l_1))
(\psi_{Lf_2}^\dagger(k_2)\psi_L^{g_2}(l_2))\right.
\]
\beq
\left. +\frac{1}{8N_c}
(\psi_{Lf_1}^\dagger(k_1)\sigma_{\mu\nu}\psi_L^{g_1}(l_1))
(\psi_{Lf_2}^\dagger(k_2)\sigma_{\mu\nu}\psi_L^{g_2}(l_2))\right]
\la{Y2}\eeq
For the antiinstanton-induced vertex $Y^{(-)}$ one has to replace
left handed components by right-handed. In \eq{Y2} I have included
the factor $\lambda_+$ and, moreover, fixed it from the saddle-point
equation. With the normalization of \eq{Y2} the dynamical mass $M(k)$
satisfies exactly the same gap equation \ur{gap} as in the case of
$N_f=1$. Note that the second (tensor) term is negligible at large
$N_c$. Using the identity

\beq
2\epsilon^{f_1f_2}\epsilon_{g_1g_2}=\delta_{g_1}^{f_1}\delta_{g_2}^{f_2}
-(\tau^A)_{g_1}^{f_1}(\tau^A)_{g_2}^{f_2}
\la{id}\eeq
one can rewrite the leading (first) term of \eq{Y2} as

\beq
(\psi^\dagger\psi)^2+(\psi^\dagger\gamma_5\psi)^2-(\psi^\dagger\tau^A\psi)^2
-(\psi^\dagger\tau^A\gamma_5\psi)^2
\la{NJLf}\eeq
which resembles closely the Nambu--Jona-Lasinio model. It should be
stressed though that in contrast to that {\em at hoc} model the
interaction \ur{Y2} {\it i}) violates explicitly the $U_A(1)$ symmetry,
{\it ii}) has a fixed interaction strength and {\it iii}) contains an
intrinsic ultraviolet cutoff due to the formfactor function $M(k)$.
This model is known to lead to chiral symmetry breaking, at least at
large $N_c$ when the use of the mean field approximation to the model
is theoretically justified.

\vspace{.5cm}
\underline{Any $N_f$}

At arbitrary $N_f$ the {\em leading} term at $N_c\rightarrow\infty$
can be written as a determinant of a $N_f\times N_f$ matrices composed
of quark bilinears:

\[
Y_{N_f}^{(\pm)}\stackrel{N_c\rightarrow\infty}{=}\left(\frac{2V}{N}\right)^
{N_f-1}\int d^4x\;{\rm det}_{N_f}J^{(\pm)},
\]
\beq
J_{fg}^{(\pm)}(x)=\int\frac{d^4kd^4l}{(2\pi)^8}e^{i(k-l,x)}
\sqrt{M(k)M(l)}\psi_f^\dagger(k)\frac{1\pm\gamma_5}{2}\psi^g(l).
\la{YNf}\eeq
Again, one can prove that at least at large $N_c$, this interaction
leads to the spontaneous chiral symmetry breaking, with the dynamical
mass determined by the gap equation \ur{gap}, and the chiral condensate
given by \eq{ccn}. The bosonization of these interactions has been
performed in ref. \cite{DP3}; it paves the way to studying
analytically various correlation functions in the instanton vacuum.

A separate issue is the application of these ideas to hadrons made
of heavy quarks \cite{DPP1} and of light and heavy ones \cite{CNZ}.

\section{QCD at still lower energies}
\setcounter{equation}{0}
\def\theequation{\arabic{section}.\arabic{equation}}

Using the packing fraction of instantons $\bar\rho/\bar R\simeq 1/3$ as
a new algebraic parameter one observes that all degrees of freedom in
QCD can be divided into two categories: {\it i}) those with masses $\ge
1/\bar\rho$ and {\it ii}) those with masses $\ll 1/\bar\rho$. If one
restricts oneself to low-energy strong interactions such that
momenta are $\ll 1/\bar\rho\simeq 600\;MeV$, one can neglect the
former and concentrate on the latter. There are just two kind of
degrees of freedom whose mass is much less than the inverse average
size of instantons: the (pseudo) Goldstone pseudoscalar
mesons and the quarks themselves which obtain a dynamically-generated
mass $M\sim (1/\bar\rho)(\bar\rho^2/\bar R^2)\ll 1/\bar\rho$. Thus in
the domain of momenta $k\ll 1/\bar\rho\;$ QCD reduces to a remarkably
simple though nontrivial theory of massive quarks interacting with
massless or nearly massless pions. It is given by the partition
function \cite{DP2,DP3}

\beq
{\cal Z}_{QCD}^{{\rm low\; mom.}}=\int D\psi D\psi^\dagger
\exp\left[\int d^4x
\psi^\dagger\left(i\!\not\partial+iMe^{i\pi^A\tau^A\gamma_5}\right)
\psi\right].
\la{lowmom}\eeq
Notice that there is no kinetic energy term for the pions, and that
the theory is not a renormalizable one. The last circumstance is due to
the fact that it is an effective low-energy theory; the ultraviolet
cutoff is actually $1/\bar\rho$.

There is a close analogy with solid state physics here. The microscopic
theory of solid states is QED: it manages to break spontaneously the
translational symmetry, so that a Goldstone excitation emerges, called the
phonon. Electrons obtain a "dynamical mass" $m^*$ due to hopping from
one atom in a lattice to another. The "low energy" limit of solid
state physics is described by interactions of dressed electrons
with Goldstone phonons.  These interactions are more or less fixed by
symmetry considerations apart from a few constants which can be deduced
from experiments or calculated approximately from the underlying QED.
Little is left of the complicated dynamics at the atom scale.

What Petrov and I have attempted, is a similar path: one starts with
the fundamental QCD, finds that instantons stabilize at a relatively low
density and that they break chiral symmetry; what is left at low
momenta are just the dynamically massive quarks and massless pions.
One needs two scales to describe strong interactions at low momenta:
the ultra-violet cutoff, whose role is played by the inverse instanton
size, and the dynamical quark mass proportional to the square root
of the instanton density. If one does not believe our variational
calculations of these quantities one can take them from experiment.

If one integrates off the quark fields in \eq{lowmom} one gets the
{\em effective chiral lagrangian}:

\[
S_{eff}[\pi]=-N_c{\rm Tr\;ln}\left(i\!\not\partial+iMU^{\gamma_5}\right),
\]
\beq
U=\exp(i\pi^A\tau^A),\;\;\;U^{\gamma_5}=\exp(i\pi^A\tau^A\gamma_5),
\;\;\;L_\mu=iU^\dagger\partial_\mu U.
\la{chilagr}\eeq

One can expand \eq{chilagr} in powers of the derivatives of the pion
field and get \cite{DP2,DP3}:

\[
S_{eff}[\pi]=\frac{F_\pi^2}{4}\int d^4x\;{\rm Tr}L_\mu^2
-\frac{N_c}{192\pi^2}\int d^4x \left[2{\rm Tr}(\partial_\mu L_\mu)^2
+{\rm Tr}L_\mu L_\nu L_\mu L_\nu\right]
\]
\beq
+\frac{N_c}{240\pi^2}\int
d^5x\;\epsilon_{\alpha\beta\gamma\delta\epsilon}\;
{\rm Tr}L_\alpha L_\beta L_\gamma L_\delta L_\epsilon +...
\la{derexpan}\eeq
The first term here is the old Weinberg chiral lagrangian with

\beq
F_\pi^2=4N_c\int\frac{d^4k}{(2\pi)^4}\frac{M^2(k)}{[k^2+M^2(k)]^2};
\la{Fpieq}\eeq
the second term are the four-derivative Gasser--Leutwyler terms
(with coefficients which turn out to agree with those following from
the analysis of the data); the last term in \eq{derexpan} is the
so-called Wess--Zumino term.  Note that the $F_\pi$ constant diverges
logarithmically at large momenta; the integral is cut by the momentum-
dependent mass at $k\sim 1/\bar\rho$, so that one gets the same
expression as in a different approach described in section 10, see
\eq{Fpi}.

An ideal field of application of the low-momentum partition function
\ur{lowmom} is the quark-soliton model of nucleons \cite{DPP2} --
actually the model has been derived from this partition function. The
size of the nucleon $\sim (250\; MeV)^{-1}$ is much larger
than the size of instantons $\sim (600\; MeV)^{-1}$; hence the
low-momentum theory \ur{lowmom} seems to be justified. Indeed, the
computed static characteristics of baryons like formfactors, magnetic
moments, etc., are in a good accordance with the data (for a review see
ref. \cite{G}).  What is not yet computed, are the nucleon structure
functions at a low normalization point, however given the previous
experience, I can bet it would go through the data points.

\section{How instantons may help confinement}
\setcounter{equation}{0}
\def\theequation{\arabic{section}.\arabic{equation}}

Our analytical calculations sketched in these lectures, the extensive
numerical studies of the instanton vacuum by Shuryak and collaborators
and the recent direct lattice measurements -- all point out that
instantons play a crucial role in determining the world of light
hadrons, including the nucleon. Confinement has not much to do with it
-- contrary to what has been a common wisdom a decade ago and in what
many people still believe. Nevertheless, confinement is a property of
QCD, and one needs to understand the confinement mechanism. What can
be said today is that confinement must be "soft": it should destroy
neither the successes of the perturbative description of high-energy
processes (no "string effects" there) nor the successes of instantons
at low momenta. At the moment I can think of two possible scenarios
\footnote{This section is based on our speculations with Victor Petrov.
The reader is kindly asked not to be too severe to these qualitative
considerations}:

\vspace{.3cm}
A) Instantons have a micro-structure, like merons \cite{CDG};

B) Confinement is due to monopoles which are massless because of
instantons.

\vspace{.3cm}
Let me stress that, contrary to the case of matter where objects
exist by themselves, in field theory of the vacuum one has first of
all to create the objects (like monopoles) which could bring in
confinement. Therefore, only such objects can give a sizeable effect
whose mass is effectively zero, so that one does not loose energy to
create them. Note that in a sense instantons have zero mass since they
have finite action, and action is mass times the (infinite) observation
time. Merons like instantons have also finite action (if one takes care
to cut them both at large and small distances), therefore there is
nothing wrong in principle with merons. The only field-theoretical
example of confinement we know today is the famous Polyakov's
example in the 2+1 dimensional Georgi--Glashow model \cite{Pol}, and
it is an example of a meron type \cite{D3}. Moreover, a meron pair
resembles one instanton \cite{CDG}, so a confinement mechanism based
on merons might fit in well into the successful instanton vacuum.

However, there is a more popular version of confinement on the
market, as due to the Mandelstam--'t Hooft monopole condensation
mechanism. According to this mechanism, monopole-like {\em particles}
are somehow formed out of the gluon fields, they develop in time,
interact and annihilate in pairs, and the crucial thing about them is
that they form a quantum-mechanical condensate characterized by a
macroscopic wave function -- like the Cooper pairs of electrons in the
superconductor. The confinement of colour electric charges is then due
to the dual Meissner effect.

Contrary to the case of electrons which exist in abundance in any
sample so that one needs just a small attraction to bind them into a
condensate, monopoles have to be first of all created, and that costs
energy. Therefore, this mechanism has a chance only if the mass of
the monopoles is effectively zero. However the monopole mass can be
estimated as ${\cal M}_{mon}\sim 4\pi/g^2\cdot(inverse\;size)$.
"$1/g^2$" appears here because the monopole should carry a unit
magnetic flux.  Therefore, in the weak coupling regime there is no
chance for monopoles to condense. This is a point where instantons may
help.

Let us write down the effective action for gluons in the background
field of one instanton:

\beq
{\cal Z}=\int DA_\mu \exp\left(-\int \frac{F_{\mu\nu}^2}{4g^2}\right)
\int d^4z\:d\rho\:dO\:d_0(\rho)\exp\left(-\frac{2\pi^2\rho^2}{g^2}
F_{\mu\nu}^a(z)O^{ab}\thb{\mu}{\nu}{b}+...\right).
\la{effact1}\eeq
This effective action has been first suggested in ref. \cite{CDG} to
describe the leading dipole-dipole interactions of instantons, then it
has been re-derived in a more general form in ref. \cite{VZNS}. Later
on Yung showed \cite{Y} that its domain of applicability is wider than
anticipated; recently this effective action has been used to derive the
\II interaction potential up to the next-to-next-to leading order \cite
{DPol}. This effective action reproduces also the instanton field
itself as an expension in $\rho^2/(x-z)^2$. For an anti-instanton one
has to replace $\bar\eta\rightarrow\eta$.

For a grand canonical ensemble of \IIs one writes an effective action

\[
{\cal Z}^{I\bar I}=\sum_{N_{+,-}}\frac{1}{N_+!}\frac{1}{N_-!}
\left(\int d^4z...\exp(...\bar\eta)\right)^{N_+}
\left(\int d^4z...\exp(...\eta)\right)^{N_-}
\]
\beq
=\int DA_\mu \exp\left\{-\int \frac{F_{\mu\nu}^2}{4g^2}-
\int d^4z\:d\rho\:dO\:d_0(\rho)\left[
e^{-\frac{2\pi^2\rho^2}{g^2}(FO\bar\eta)}+
e^{-\frac{2\pi^2\rho^2}{g^2}(FO\eta)}\right]
\right\}.
\la{effactN}\eeq

If one expands the exponents (in the exponent) in powers of
$F_{\mu\nu}$ one observes that the linear term is zero owing to
integrations over orientations, and the quadratic term actually
corresponds to the renormalization of the gauge coupling due to
the instanton medium:

\[
\frac{1}{4g^2}\rightarrow\frac{1}{4g^{*2}}
=\frac{1}{4g^2}-\frac{1}{4(N_c^2-1)g^4}\int d\rho\;d(\rho)(2\pi\rho)^4
\]
\beq
=\frac{1}{4g^2}\left(1-\frac{(2\pi)^4}{g^2(N_c^2-1)}
\frac{N}{V}{\bar\rho^4}\right).
\la{rencou}\eeq

We see that if the packing fraction is large enough the effective
coupling $g^{*2}$ for long range fields blows up. It means that if
the instanton packing is close to some critical value, one does not
need much energy to create a monopole in such medium; that is the
necessary condition for their condensation. To get an accurate estimate
of the critical density is not so easy, though. To that end one needs
to have a good understanding of the usual perturbative renormalization
of the charge by instantons: what is the precise argument of the
running coupling constant $g^2$ in \eq{rencou}? Using the numbers
obtained in ref. \cite{DP1} we get that the density is about half that
of the critical, but the uncertainty of this estimation is high: it
could be close to the critical as well.

Closing this section, I would like to mention that the monopole
condensation of Mandelstam and 't Hooft is probably not what we in fact
need -- the confinement mechanism should be probably more subtle.
The essence of that mechanism is the Landau--Ginzburg or the Higgs
effect --- but for dual Yang--Mills potentials. 't Hooft has elaborated
it in some detail \cite{tH2}: all fields are classified in respect to
the maximal abelian $U(1)\times U(1)$ subgroup of the $SU(3)$ colour
group, and monopoles have magnetic charges in respect to those $U(1)$
subgroups. If they condense all particles which carry electric charges
in respect to those $U(1)$ subgroups are confined. Particles which
happen to be neutral are not, though. For example, two gluons out
of eight are "photons" of these $U(1)$ subgroups, so they are neutral
and are not confined, instead they may obtain a "magnetic" mass which
is of the order of the string tension, that is about $420\;MeV$, maybe
up to a factor of 2 heavier.
Probably such objects should show up as resonances in usual particle
production, but we do not know of two such additional states. Even
worse, quark- antiquark pairs belonging to the colour octet
representation but having colour $T_3$ and $Y$ zero are also neutral in
respect to the $U(1)$ subgroups, so they should also exist and be
observable. There are two such additional states for each set of meson
quantum numbers. There would be also five additional types of baryon states 
which are not colour singlets but which are neutral in respect to the both
$U(1)$ subgroups. And of course nothing prevents monopoles themselves
from getting into an experimentalist's detector, if only they do not,
in addition, carry electric charges in respect to the $U(1)$ subgroups.

Therefore, I think that what we actually need in QCD is not
condensation of monopoles but rather a {\em pre}-condensation,
something of the kind of the Berezinsky--Kosterlitz--Thouless phase,
characterized by large anomalous dimensions of the mono\-pole (and
probably also gluon) fields. To obtain that one also needs massless
or effectively massless monopoles, and that is where instantons might
help.

\vspace{.5cm}
{\bf Acknowledgements}. These lectures have been written up while
visiting the European Centre for Theoretical Studies (ECT*) in Trento.
I acknowledge the support of the ECT* and of the I.N.F.N.
My special gratitude is to Victor Petrov with whom we worked together
for many years on the topics discussed in these lectures.

\end{document}